\documentclass{MIPRO}

\usepackage{cite}
\usepackage{amsmath,amssymb,amsfonts}
\usepackage{algorithm}
\usepackage{algorithmicx}
\usepackage[scientific-notation=true]{siunitx}
\usepackage{algpseudocode}
\usepackage{graphicx}
\usepackage{textcomp}
\usepackage{xcolor}
\usepackage[T1]{fontenc}
\usepackage[ancient]{flushend}
\usepackage{multirow}
\usepackage{multicol}
\usepackage{array}
\usepackage{xurl}
\usepackage{hyperref}

\begin{document}

\title{Batched matrix operations on distributed GPUs with application in theoretical physics}

\author{
\IEEEauthorblockN{
Nenad Miji\'{c}\IEEEauthorrefmark{1},
Davor Davidovi\'{c}\IEEEauthorrefmark{1}
}

\IEEEauthorblockA{\IEEEauthorrefmark{1} %1st affiliation 
Centre for Informatics and Computing, Ru\dj er Bo\v skovi\' c Institute, Zagreb, Croatia, \{nmijic, ddavid\}@irb.hr}

\{nenad.mijic, davor.davidovic\}@irb.hr

}

\maketitle

\begin{abstract}
One of the most important and commonly used operations in many linear algebra functions is matrix-matrix multiplication (GEMM), which is also a key component in obtaining high performance of many scientific codes. It is a computationally intensive function requiring $O(n^3)$ operations, and its high computational intensity makes it well-suited to be significantly accelerated with GPUs. Today, many research problems require solving a very large number of relatively small GEMM operations that cannot utilise the entire GPU. To overcome this bottleneck, special functions have been developed that pack several GEMM operations into one and then compute them simultaneously on a GPU, which is called a batch operation. In this research work, we have proposed a different approach based on linking multiple GEMM operations to MPI ranks and then binding multiple MPI ranks to a single GPU. To increase GPU utilisation, more MPI ranks (i.e. GEMM operations) are added. We implement and test this approach in the field of theoretical physics to compute entanglement properties through simulated annealing Monte Carlo simulation of quantum spin chains. For the specific use case, we were able to simulate a much larger spin system and achieve a speed-up of up to $35\times$ compared to the parallel CPU-only version.
\end{abstract}

\renewcommand\IEEEkeywordsname{Keywords}

% Insert sections
\section{Introduction}\label{sec:intro}

Matrix multiplications are the fundamental building blocks in many functions of linear algebra as well as in a variety of scientific and industrial codes. The key kernel of linear algebra responsible for the high performance of many computer architectures is matrix-matrix multiplication (GEMM). It is usually one of the first kernels to be optimised and adapted for different processors and computer architectures such as multi-core CPUs, Intel Xeon Phi, ARM or GPUs~\cite{Anderson1999LAPACKGuide, Haidar2015MAGMAComputing, Guney2017OptimizingArchitecture}.
Matrix multiplication is a computationally intensive problem whose performance is limited by processor speed rather than memory bandwidth and latency. The arithmetic intensity (the ratio of floating point operations per data transferred) is defined by the number of rows or columns and, for sufficiently large matrices, can reach a performance close to the theoretical peak performance of the computing system.
Due to its high arithmetic intensity, matrix multiplication is well suited to efficiently hide slow data movement from slower memory locations (e.g., disc or main memory) to faster, near-processor memory (e.g., cache memory or global and shared memory on GPUs), but careful implementation of matrix multiplication is required~\cite{Goto2008}.
Highly optimised, fine-tuned implementations of matrix multiplication can be found in numerous numerical computation libraries for multi-threaded and multi CPU systems (BLAS ~\cite{Dongarra1988AnSubprograms}, LAPACK~\cite{Anderson1999LAPACKGuide}, OpenBLAS~\cite{OpenBLASLibrary}, MKL~\cite{Intel_OneAPI_MKL}), GPU-accelerated and heterogeneous platforms (cuBLAS~\cite{cuBLAS}, MAGMA ~\cite{Haidar2015MAGMAComputing, TheUniversityofTenneesse2014MAGMAArchitectures}) and distributed memory systems (ScaLAPACK~\cite{Choi2003ScaLAPACK:Computers, Choi1996}).
%From the above it is clear that the matrix multiplication would be able to achieve a close to the theoretical peak performance of the computing system, however a careful implementation is required~\cite{Goto2008} in order to achieve the highest performance.

Although very high performance and resource utilisation can be achieved in matrix multiplication, when working with small matrices the impact of overlapping data transfers with useful computations can be significantly reduced. This shortcoming becomes even more apparent when the computation is moved to the GPU devices, as the matrix is usually not large enough to fully utilise the entire GPU. In addition, for each multiplication, the data must be transferred between the main memory and the GPU memory via a relatively slow CPU-GPU interconnection, which is characterised by high latency.

The problem described above becomes even more challenging when solving problems involving a large number of small matrix multiplications. In these problems, the transfer of a large number of small matrices to GPUs can easily become a time-dominant part of the overall execution due to the latency issues. An example where a large number of small matrix computations are required are domain decompositions~\cite{Agullo2015OnSolvers}, 3D graphics transformations in the Level 3 Cascading Style Sheets specification in web browsers~\cite{Higham2016}, Astrophysics~\cite{Messer2012MulticoreCode}, Finite Element Methods~\cite{Abdelfattah2016High-performanceGPUs} and in Machine Learning and Artificial Intelligence~\cite{Chetlur2014CuDNN:Learning, Abdelfattah2020MatrixPrecisions, Georganas2019High-PerformanceBlock}. 

To solve these types of problems efficiently on GPUs, the common approach is to group or package many small matrix operations into one larger operation, which is then transferred and processed simultaneously on the GPU. By packing multiple operations into one larger operation, the number of memory transfers can be greatly reduced, which decreases latency and increases GPU utilisation. This type of operation is commonly referred to as {\tt batched}matrix operations.

In this paper, we present an alternative approach to batched matrix operations (called \textit{batchedGemm} in the rest of this research) with our own approach based on scheduling multiple MPIs on a single GPU. The original contribution of this research are the following:

\begin{itemize}
    \item Implicitly packing a larger number of small matrices on a single GPU using MPI ranks,
    \item The proposed model allows easy scaling to a larger number of GPUs and across compute nodes.
    \item Improved performance compared to the state-of-the-art Numpy batched approach (up to $22\%)$. 
    \item Achieved $35\times$ speedup compared to the CPU-only version on the same amount of resources for the use case Simulated Annealing Monte Carlo Simulation.
%\item Use case has been built entire in Python programming language thus can be easily adapted and used in combination with a variety of Python libraries.
\end{itemize}

The rest of the paper is organized as follows. In Section \ref{sec:batchedGemm} a brief introduction to problem of computing many matrix operations is given, with state-of-the art methods and libraries used to solve them. Our approach is solving batch GEMM operations is presented in Section \ref{sec:alg} together with the targeted use-case. The numerical results and the achieved performance are discussed in Section \ref{sec:results}. The final notes and the conclusion of our work is given in Section \ref{sec:conc}   

\section{Batched GEMM and related work}\label{sec:batchedGemm}

The aim of this research is to find an efficient solution for many small matrix-matrix multiplications on the GPU. A set of matrix multiplications can be defined as follows:

\begin{equation}\label{eq:gemm} C_i = \alpha A_i B_i + \beta C_i, \quad i = 1,\ldots N
\end{equation}

where $A_i \in \mathbb{C}^{m,k}$, $B_i\in \mathbb{C}^{k,n}$, $C_i\in \mathbb{C}^{m,n}$ are complex matrices and $N$ is the number of matrix-matrix multiplications to be calculated. 
The common approach to compute a single matrix is to partition (tile) $C$ into multiple tiles, each processed independently by a block of threads on the GPU, with each thread computing one element of the matrix $C$. Parallelism can be exploited between tiles, i.e. several tiles run simultaneously, and within tiles, i.e. each tile is processed by many threads. When $C$ is large, this approach can attain close to the peak performance of the GPU, as there are enough tiles to fully utilise the symmetric multiprocessors of the GPU. However, when $C$ is small, the number of tiles is insufficient to fully utilise the GPU. Processing a large number of small GEMMs requires a large number of data copies between main memory and GPU memory, resulting in the application being memory-bound rather than compute-bound.%This approach might be a good solution on modern processors (CPUs) with complex cache hierarchy, but attains very low performance on the GPU. The main drawbacks are the latency, as a result of a large number of data copies to the GPU memory and low utilization of the GPU shared multiprocessors caused matrices too small to occupy all bandwidth and computational resources on a device. 

In order to exploit the full power of GPU devices, an approach has been developed based on packing many small matrix operations into a larger operation called a batch. The basis for the idea of batch matrix multiplication is the tiling algorithm, which divides the matrix into a 2D grid of tiles, each of which represents a subproblem that is then processed simultaneously on the GPU. The initial matrix is transferred once (a large memory copy), while the division into smaller subproblems is done on the device~\cite{Agullo2009NumericalProjects}. Performance can be fine-tuned by adjusting the tile size depending on the available resources on the GPU device. The implementation of this approach can be found in MAGMA ~\cite{TheUniversityofTenneesse2014MAGMAArchitectures, Haidar2015MAGMAComputing} dense linear algebra library. One of the main drawbacks of the latter approach is the fixed size of the matrices, e.g. all matrices of $A$ should have the same size and be stored in contiguous memory locations. To overcome this problem, a variable-size GEMM was proposed~\cite{Abdelfattah2016PerformanceGPUs, Abdelfattah2017NovelGPUs}, which divides each GEMM (matrices $A$, $B$ and $C$, Eq.~\ref{eq:gemm}) into tiles processed by thread blocks. Then each $C$ can be processed by its 2D grid of thread blocks. 
In the paper~\cite{Masliah2016High-PerformanceMatrices}, the authors propose a special kernel for matrices with size less than 32, which are often encountered in Big Data analytics, machine learning and high-order finite element method (FEM). The approach is based on aggregating multiple thread blocks to compute two or more tiny matrices $C$ (matrices smaller than the warp size) and achieves performance within 90\% of a large single GEMM.

In batch computation by tiling and batching, the size of the tiles and the batch size, respectively, are the two most important performance parameters. Although different matrix sizes are possible, the tiling strategy is the same, which is not the best solution for different matrix sizes. To overcome this problem, a novel solution is proposed~\cite{Li2019AGPUs} that allows different tiling strategies (different tile sizes for different $C$) and batch strategies (the number of tiles allocated to a thread block). The proposed work achieves $1.23\times$ speedup on the GoogleNet case study compared to the state-of-the-art CUBLAS and MAGMA implementations.

In some research areas, such as latency-critical deep neural networks consisting of a large number of small batch sizes, the classical batch approach provides very limited GPU usage. Instead of packing smaller operations into a batch, GPU resource usage can be split across multiple processes or tasks. In~\cite{Jain2018DynamicInference}, the authors experimented with two approaches. The first is to put each process in a separate CUDA context and then use a software scheduler to interleave the executions on the GPU. The second approach is to use the NVIDIA Multi-Process Service (MPS)~\cite{Multi-ProcessDocumentation} server to partition multiple processes into separate queries across a pool of CUDA streams. Recent research \cite{OroutzoglouExplorationWorkloads} shows that computing a number of $2k\times 2k$ and $8k\times 8k$ matrix multiplications with MPS can achieve a speedup of up to $4.5\times$ compared to the native CUDA scheduler.
%\section{Algorithmic approach}
\section{Task-based batch GEMM}
\label{sec:alg}

In this research we tackle a problem of processing a number of embarrassingly parallel processes, each computing a series of small matrix operations, each operation not large enough to fully utilize the GPU. Because of its embarrass parallelism, there is no need for an explicit synchronization between the processes thus any packing of many operations into one batch operations will cause an unnecessary synchronization points in the execution. 

\begin{figure}
  \centering
  \includegraphics[scale=0.5]{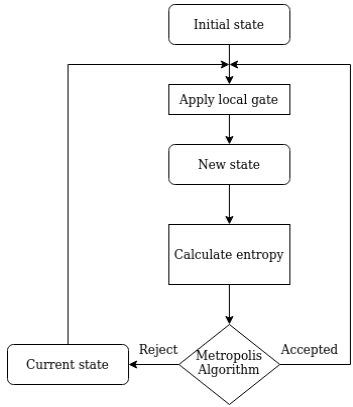}
  \caption{Simulation model.}
  \label{fig:pseudo}
\end{figure}

An example is the simulated annealing Monte Carlo (MC)~\cite{Vanderbilt1984AVariables} simulation of frustrated one-dimensional transverse-field Ising model~\cite{Cipra2018AnModel} with $S$ spins, Fig.~\ref{fig:pseudo}. The simulation starts with the initial calculating of the ground state. The main body of the simulation consists of applying local unitary operations (Fig.~\ref{fig:pseudo}, \texttt{Apply Local Gate}) on the neighbouring pair of spins in the lattice, chosen at random, and computing the new entanglement (Fig.~\ref{fig:pseudo}, \texttt{Calculate Entropy}). Finally, the decision is made to accept this new state (Metropolis algorithm, Fig.~\ref{fig:pseudo} \texttt{M.A}) or keep the current state. Then, repeat the process with the unitary operator on the chosen state.

The main computational bottleneck is the \texttt{Calculate Entropy} in which two complex matrices with dimensions depending on the number of spins $\left( M = N = 2^{S - \lfloor S/2 \rfloor}, K = 2^{\lfloor S/2 \rfloor} \right)$ are multiplied. Since the next step of the MC simulation depends on the state computed in the current step, the MC procedure is an inherently sequential code and the loop over MC steps (Alg.~\ref{alg:pseudo}, line \ref{alg:pseudo:steps}) cannot be parallelised. Therefore, only matrix multiplications (line~\ref{alg:pseudo:gemm}) can be parallelised, but only one GEMM at the time can be processed (due to inter-step dependencies). The number of MC steps ($N_s$) (line~\ref{alg:pseudo:steps}) can be arbitrary large, but in practice, this number can be counted in millions, thus limiting the performance gain especially when GEMM operations are small. 

To increase the statistical significance of the simulation results, a large number of repetitions ($N_p$) of the Monte Carlo procedure are required (line~\ref{alg:pseudo:procedure}). The MC procedures are mutually independent and no data transfer or communication between them is necessary, making the problem an embarrassingly parallel one and straightforward to parallelise. A typical real-world use-case with $S = 21$ spins, $100$ MC procedures and $10^6$ steps each, will require to compute a series of GEMM operations of size $\left( M = N =1024, K = 2048 \right)$ in each MC procedure.

\begin{algorithm}[t]
  \caption{Algorithm: Tasked based GEMM}\label{alg:pseudo}
  \begin{algorithmic}[1]
    \Require {Number of spins $S$}
    \Ensure{Average entropy}
    \State Compute initial state \label{alg:pseudo:ground}
    \For{$i \in [1\cdots N_p]$} \label{alg:pseudo:procedure}
    \Comment{Loop over MC procedures}
        \For{$s \in [1\cdots N_s]$} \label{alg:pseudo:steps}
        \Comment{Loop over MC steps}
            \State Apply unitary operator \label{alg:pseudo:localGate}
            \Comment{Apply Local Gate}
            \State Generate $A, B$ \label{alg:pseudo:gen}
            \State $C \rightarrow GEMM(A,B)$ \label{alg:pseudo:gemm}
            \Comment{Calculate Entropy}
        \EndFor
    \EndFor
    \State Compute average entropy
  \end{algorithmic}
\end{algorithm}

To efficiently solve the above problem, in this research we proposed a solution based on the spatial sharing of the GPU resources among multiple processes and tasks. Within the MC procedures the matrix multiplications are offloaded to the GPU (Alg.~\ref{alg:pseudo}, line~\ref{alg:pseudo:gemm}), while MC procedures (line \ref{alg:pseudo:procedure}) are distributed using Message Passing Interface (MPI), each MPI rank calculating one MC procedure independently. Multiple MPI processes (MC procedures) are assigned to same device in order to saturate it with other ready to execute kernels. This can achieve better utilization at the expense of less speed up per single kernel.
This approach is a simple from the programming point of view, because it leaves to the GPU internal scheduler to allocate unused resources for other kernels (processes), a kind of tasked based parallelism of independent simulation procedures. The proposed approach with multiple task utilizing the same GPU has a higher potential of improving throughput and lowering latency, because multiple memory transfer operations and kernels can be arbitrarily and asynchronously overlapped.   

The drawback of this approach is that each process creates its unique context on the device that leads to the process synchronization, since exclusively only one context can be active on the device at the time. To overcome this problem NVIDIA has devised Multi Processing Service (MPS), a software layer between CUDA driver and multi process program, which creates single context and routes all CUDA calls/kernels through it. Starting with the Volta architecture, each process sends jobs to the device separately without passing through the MPS server where each process is scheduled in its own queue. It is advisable to use the MPS when GPU utilization per process is low and want to achieve multiple kernels to execute concurrency.
\section{Results}\label{sec:results}

In this section, we present and analyse the performance of our approach in real use cases and compare it with the parallel CPU -only implementation and the GPU-accelerated batch implementation. In the rest of the paper, we refer to our approach as {\tt taskedGEMM} and the classical batch approach as {\tt batchedGEMM}. In both cases, we test how the size of the batch (or in the case of {\tt taskedGEMM} the number of MPI ranks per GPU) affects the performance and how it scales with the addition of more procedures running in parallel for different problem sizes. 

The tests were conducted on the GPU partitions of the HPC Vega supercomputer at the Institute of Information Science, Slovenia. 
Each node is equipped with two 64-core AMD Rome 7H12 CPUs and four NVIDIA Tesla A100 GPUs with a peak performance of 19.5 TFlops 
(double precision, using Tensor Cores) per GPU.

For benchmarking our MPI-based approach, we extend the original Monte Carlo simulation code written in the Python programming language. 
The original code takes advantage of highly tuned NumPy and SciPy libraries for multithreaded matrix multiplication on shared memory systems. Although the NumPy implementation of GEMM achieves better results than the sequential code, it turns out that it cannot handle larger simulation sizes ($S > 19$) in a reasonable time. Therefore, the matrix generation and the most time-consuming parts of the code - the matrix multiplications - are offloaded to the GPUs. For the GPU-accelerated computations, we used the CuPy 10.0.0 library with CUDA 11.4.0 and cuBLAS 11.5.2.43. To distribute the workload between the compute nodes and GPUs, we use SLURM 20.11 Workload Manager and OpenMPI 4.1.1 (UCX 1.11.2).

\subsection{Single GPU benchmark}
In the first benchmark, we want to find the best binding policy of MPI ranks (i.e.~number of Monte Carlo procedures $N_p$) per GPU.
For this test, we focused on a single GPU and small problem sizes with the number of MC steps $N_s=1000$, the number of MC procedures $N_p = \{1,2,4,\ldots,16\}$ and the number of spins $S = \{15, 19, 21, 23\}$. Since there is a linear dependence between the MS steps, we can test with a smaller number of steps without loss of generality and then simply extrapolate linearly to a desired number of MC steps.

\begin{figure}[tb]
  \centering
  \includegraphics[width=\columnwidth]{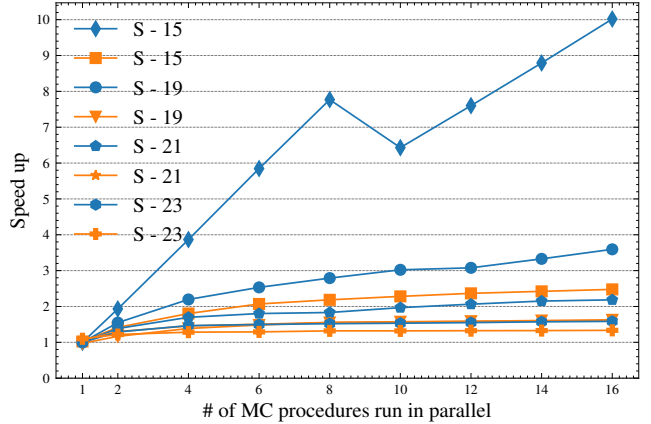}
  \caption{Speed-up of batched (orange) and tasked (blue) GEMM compared to sequential execution.}
  \label{fig:rrun}
\end{figure}

Fig.~\ref{fig:rrun} shows the speed-up of {\tt bachedGEMM} and {\tt taskedGEMM} compared to the sequential version, i.e. when MC procedures are executed individually on the GPU. The number of MPI ranks in {\tt batchedGEMM} is 1 and the batch size (number of MC procedures packed in a batch ) is the number of procedures (represented by the x-axis, Fig.~\ref{fig:rrun}). In {\tt taskedGEMM}, the number of MPI ranks is equal to the number of procedures (x-axis), since each MPI rank executes one MC procedure and all MPI ranks use the same GPU.
The figure shows that the {\tt taskedGEMM} achieves better speed-up than the batch approach in all cases. As expected, as the number of MC procedures executed in parallel increases, we observe a steady speed-up up to a certain point. For example, in {\tt taskedGEMM} with $15$ spins, the speed-up is linear up to 8 MC procedures. This is because the matrices involved in GEMM are too small ($256 \times 128$) to make full use of the test GPU. When more MC procedures are added, the speed-up slows down as the GPU is overused, causing some MCs to end up waiting for GPU resources. For a larger spin system, e.g. $S = 23$, the over-utilisation of GPU resources occurs even earlier, with only $4$ MC procedures, as the matrices involved in the computations are much larger ($4094\times2048$). The higher speed-up of {\tt taskedGEMM} indicates that interleaving GEMM operations leads to a slightly better resource utilisation and thus to higher efficiency.

%\begin{figure}
%  \centering
%  \includegraphics[scale=0.3]{img/relative_runtime_tb.pdf}
%  \caption{Speed up of tasked implementation over batched.}
%  \label{fig:rrun_tb}
%\end{figure}

\begin{figure}[tb]
  \centering
  \includegraphics[width=\columnwidth]{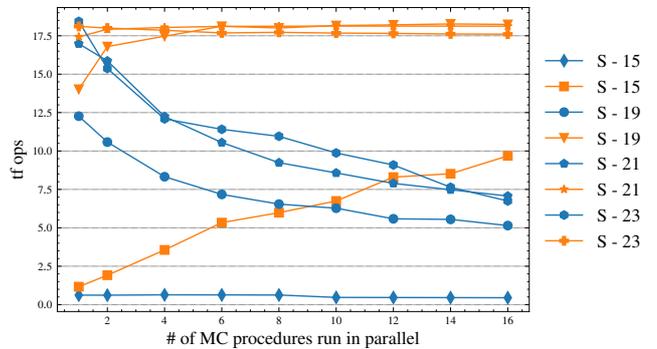}
  \caption{Performance of the single GEMM operation as a function of the number of MC procedures for batch (orange) and task (blue) variants. }
  \label{fig:tflops}
\end{figure}

The average performance (in TFLOPS/s) achieved per GEMM operation is given in Fig.~\ref{fig:tflops}. To calculate the performance of a single GEMM in {\tt batchedGEMM}, the total performance is divided by the size of the batch. To allow a fair comparison, in the task-based model each GEMM call is counted separately and the average performance is calculated. The constant performance of {\tt batchedGEMM} for sizes $N_s=19, 21, 23$ indicates that the matrices are large enough to fully utilise the GPU even if only 1 MC procedure is computed, and that adding more matrices to the batch is not expected to increase performance further. In contrast, for $S=15,19$ batched operations, performance increases when the size of the back is increased. This shows that the size of the computational problem is still too small to fully utilise the GPU, so more operations should be added.
%As expected, by batching a number of operations the throughput can be improved. However, as free lunch theorem states that better performance can be utilized only up to the full utilization of device resources. 
%For $S = 15$ shows slight improvement in executing batched operations.

In task-based GEMM operations, there is a constant drop in performance per GEMM call when more MC procedures are added (Fig.~\ref{fig:tflops}). Performance drops significantly as the number of MC procedures increases. The reason for this behaviour is that interleaving of numerous operations (kernels) leads to an overuse of GPU resources. If there are not enough GPU resources available for a particular kernel (e.g.~GEMM function call), it remains in the queue in the CUDA scheduler. As soon as the resources are released, the kernel starts executing. Such use of GPU resources consequently leads to fluctuations in the performance of individual executions, which manifest themselves as performance loss (see {\tt taskedGEMM} and $S=19,21,23$ in Fig.~\ref{fig:tflops}). 

\begin{figure}[tb]
  \centering
  \includegraphics[width=\columnwidth]{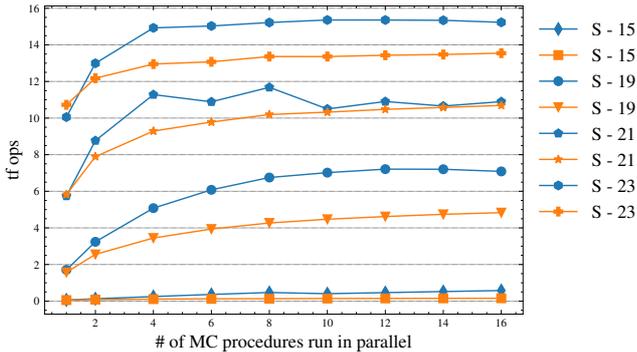}
  \caption{The performance in TFLOPS/s of the entire simulation as a function of the number of MC procedures for batch (orange) and task (blue).
  %As expected, bigger the matrix size, less parallel simulations needs to achieve better GPU utilization.
  }
  \label{fig:tflops_tot}
\end{figure}

Although the time required to compute a GEMM operation in {\tt taskedGEMM} is slower than in {\tt batchedGEMM}, the overall performance of the whole simulation is shown to be much better for the task-based approach, Fig.~\ref{fig:tflops_tot}. 
This is because interleaving different kernels (operations) on a GPU can better hide latency, reduce GPU idle time and increase utilisation. So in our use case, interleaving many independent operations has a much greater impact on performance than combining many operations into one larger but faster operation.

\subsection{Multi-GPU benchmark}

In order to run the tests on multiple GPUs, we have chosen a sufficiently large real-world use case that represents well the performance on many GPUs. For the tests, we chose a case with $S=21$ spins, $N_s=10000$ Monte Carlo steps and the fixed number of procedures $N_p=128$. Although in real-world use cases the number of MC steps is usually more than $10^6$, we can test with a smaller number of steps because the steps are interdependent and thus the time increases linearly with the number of steps. The total execution time for a larger number of steps can be easily estimated by scaling the runtime of the test with a smaller number of steps. 

The Table \ref{tab:total_time} shows the total execution time for solving the same problem with $128$ Monte Carlo procedures and different number of procedures per GPU (4, 8 and 16) with the {\tt taskedGEMM}. The test shows the results run on 1, 2, 4 and 8 GPUs. For example, the time to solve the problem with 16 tasks per GPU on only 1 GPU would be $3296$ seconds, while the time with $8$ GPUs decreases almost linearly and is $420$ seconds. The reason is that the MC procedures are independent and $16$ procedures can be processed simultaneously on one GPU, so $8$ batches have to be executed sequentially. In contrast, in the case of $8$ GPUs, all $8$ batches, each processing 16 MC procedures, can be executed simultaneously, each on its own GPU.

\begin{table}[tb]
\renewcommand{\arraystretch}{1}
\caption{Total runtime of {\tt taskedGEMM} (in seconds) for $21$ spin and $128$ MC procedures.}\label{tab:total_time}
\centering
\normalsize
\begin{tabular}{ccccc}
\multirow{2}{*}{$N_p$ per GPU}         & \multicolumn{4}{c}{Number of GPUs} \\ \cline{2-5}
         & 1  & 2 & 4 & 8 \\ \hline
4        & 3680 & 1840  & 920  & 460  \\
8        & 3392 & 1696  & 848  & 424  \\ 
16       & 3296 & 1648  & 824  & 420  \\ \hline
\end{tabular}
\vspace{0.5cm}
\end{table}

In the previous tests, we have shown that the best performance is achieved when $16$ MC procedures are executed per GPU (for a problem with $21$ spin), so we set the number of MC procedures per GPU to $16$. The total execution time and speedup of both {\tt batchedGEMM} and {\tt taskedGEMM} are shown in Table \ref{tab:speedup}. The batched and tasked GEMM variants achieve speedups of more than $26\times$ and $35\times$, respectively, compared to the CPU -only version. Note that the CPU version is a parallel code with one MC procedure per MPI rank and the GEMMs are computed using the multi-thread NumPy library. All tests were performed on 2 compute nodes, and the configuration of the CPU-only version is 64 MPI ranks per node and 2 threads per MPI rank to compute the GEMMs.

The fastest execution time we achieved on our test systems for the CPU version was achieved when the problem was solved on $32$ compute nodes with $4$ MPI ranks per node and $32$ threads per MPI rank. In this case, the total execution time for the parallel CPU variant was $3529$ seconds, noting that the result was obtained by extrapolating the computation time of 1 compute node. The speedup of {\tt batchedGEMM} and {\tt taskedGEMM} is $6.6\times$ and $8.4\times$, respectively. Although the speedup is smaller than indicated in Table \ref{tab:speedup}, the CPU-only variant requires much more computing resources to achieve this performance, exactly $16\ times$ more resources.
%To achieve the same performance using only CPUs

\begin{table}[tb]
\renewcommand{\arraystretch}{1}
\caption{ Total execution time and speed up of batch and task GEMM for $S=21$, $N_s=10000$, $N_p=128$.}
\label{tab:speedup}
\centering
\normalsize

\begin{tabular}{cccc}
         & \multicolumn{3}{c}{Algoritmic variants} \\ \cline{2-4}
         & CPU  & {\tt batchedGEMM} & {\tt taskedGEMM} \\ \hline
time & 14735 & 534      & 420     \\
speed up & 1 & 27.6      & 35.1 \\ \hline
\end{tabular}
\end{table}

%\begin{table*}[t]
%\centering
%\renewcommand{\arraystretch}{1}
%\caption{Total runtime of {\tt taskedGEMM} (in seconds) for $21$ spin and $128$ MC procedures. The values present the time required to compute.}
%\centering
%\normalsize
%\begin{tabular}{c|cccccccc}
%\multirow{2}{*}{$N_p$ per GPU}         & \multicolumn{8}{c}{Number of GPUs} \\ \cline{2-9}
%         & 1  & 2 & 3 & 4 & 5 & 6 & 7 & 8 \\ \hline
%4        &  3680  & 1840  & 1495  & 920  & 805  & 690  & 575  & 460  \\
%8        & 3392   & 1696  & 1272  & 848  & 848  & 636  & 539  & 424  \\ 
%16       & 3296 & 1648  & 1236  & 824  & 824  & 624  & 527  & 412  \\ \hline
%speed up & 1    & 27.8      & 36 
%\end{tabular}
%
%\end{table*}

\section{Conclusion}\label{sec:conc}

In this paper, we have presented a task-based model for the efficient execution of many small to medium-sized GEMM operations on GPUs. Unlike a classical batch model where many small matrix operations are packed into a larger one and executed at once on the GPU, the proposed model assigns the independent kernels/operations to different MPI ranks. Each MPI rank transfers its operations independently to the GPU, resulting in interleaving of multiple operations on the same GPU. The GPU scheduler is then responsible for the optimal distribution of the operations (kernels) on the available GPU resources. 

The model was demonstrated by simulating the frustrated one-dimensional transverse-field Ising model, a use case from theoretical physics. The model shows a significant speed-up of $22\%$ compared to the batch approach. The benchmark shows that although the batch approach has up to twice the performance per GEMM operation, our task-based approach shows higher overall performance. This shows that a model based on multiple tasks that independently offload computations to the same GPU can better utilise the GPU by running many different kernels simultaneously, reducing GPU idle time and hiding latency. 

In addition, two improvements have been made over the CPU implementation. First, the GPU code is more than $35\times$ faster than the CPU-only code and second, fewer computational resources are needed to achieve the same performance or much larger simulations can be performed in terms of number of spins, number of Monte Carlo simulations and number of steps per simulation.

\section*{acknowledgment}
This work was supported by the Croatian Science Foundation under grant number HRZZ-UIP-2020-02-4559 and the European Regional Development Fund under the grant KK.01.1.1.01.009 - DATACROSS.
The authors gratefully acknowledge the HPC RIVR consortium (www.hpc-rivr.si) and EuroHPC JU (eurohpc-ju.europa.eu) for funding this research by providing computing resources of the HPC system Vega at the Institute of Information Science (www.izum.si).

\bibliographystyle{IEEEtran}
%\bibliography{references}
%\bibliography{ms}

\end{document}